\documentclass[twoside,twocolumn,english,superscriptaddress,nofootinbib,preprintnumbers, aps]{revtex4}

\usepackage{euscript,amssymb}
\usepackage{amsthm}
\usepackage{graphicx}
\usepackage{amsfonts}
\usepackage{amsmath}
\usepackage{amssymb}
\usepackage{fancyhdr}
\usepackage{epsfig}
\usepackage{color}
\usepackage{esint}
\usepackage[unicode=true, pdfusetitle, bookmarks=true,bookmarksnumbered=false,bookmarksopen=false, breaklinks=false,pdfborder={0 0 1},backref=false,colorlinks=false]{hyperref}

\begin{document}

\title{Metric emerging to massive modes in quantum cosmological space-times}

\author{Andrea Dapor}
\email{adapor@fuw.edu.pl} \affiliation{Instytut Fizyki Teoretycznej, 
Uniwersytet Warszawski, ul. Ho\.{z}a 69, 00-681 Warszawa, Poland}

\author{Jerzy Lewandowski}
\email{jerzy.lewandowski@fuw.edu.pl} \affiliation{Instytut Fizyki Teoretycznej, 
Uniwersytet Warszawski, ul. Ho\.{z}a 69, 00-681 Warszawa, Poland}

\begin{abstract}
{We consider a massive quantum test Klein-Gordon field probing an homogeneous isotropic quantum cosmological space-time in the background. In particular, we derive a semi-classical space-time which emerges to a mode of the field. The method consists of a comparison between QFT on a quantum background and QFT on a classical curved space-time, giving rise to an emergent metric tensor (its components being computed from the equation of propagation of the quantum K-G field in the test field approximation). If the field is massless the emergent metric is of the FRW form, but if a mass term is considered it turns out that the simplest emergent metric that displays the symmetries of the system is of the Bianchi I type, deformed in the direction of propagation of the particle. This anisotropy is of a quantum nature: it is proportional to $\hbar$ and ''dresses'' the isotropic classical space-time obtained in the classical limit.}
\end{abstract}

\date{\today}

\pacs{???}

\maketitle

Quantum geometry (QG), is the idea that space-time of general relativity is just a low-energy (i.e., large-scale) effective description of gravity. At a more fundamental level, geometry is believed to be ''quantum'', and it is only because of the extreme energies that are needed to probe this level of reality that we do not observe any quantum geometry effect in the world around us. In fact, phenomenological arguments suggest that the energies at which the quantum nature of geometry cannot be disregarded are in the Planck regime ($E_\text{Pl} \approx 1.22 \times 10^{28} \ eV$, corresponding to Planck length $\ell_\text{Pl} \approx 1.62 \times 10^{-35} \ m$). As to the explicit description of quantum geometry, some proposals exist, such as Geometrodynamics \cite{GEO1, GEO2, GEO3} and Connessiodynamics \cite{CON1, CON2, CON3} (also known as Loop Quantum Gravity (LQG)). At the theoretical level, both these theories present difficulties due to the non-perturbative structure required for background-independence; nevertheless, in recent years LQG has seen a great development, and is on the verge of providing a possible complete description of the dynamics of quantum geometry. Accordingly, it is now possible to start using the theory to compute transition amplitudes, and possibly to predict observable effects which in principle can be used to falsify it.

A most successfull application of the theory is in the cosmological sector \cite{LQC1, LQC2}, in which case the common name is Loop Quantum Cosmology (LQC).\footnote{
It is to be said that LQC does not directly descend from LQG, but it is rather a finite degree of freedom model of the LQG-like quantization in the phase space of cosmology, obtained as a sector of the phase space of general relativity by restricting to homogeneous space-times.
}
Studying the dynamics of this theory, one finds that the classical singularity is removed \cite{L-FRW1, L-FRW2, L-FRW3}, being in fact replaced by a bounce, thus solving one of the greatest problems of classical general relativity (at least in the cosmological sector).

Recently,  there has been some interest in a new tool available in the context of quantum cosmological models, LQC in particular, namely that of quantum test fields probing quantum space-time. This concept was introduced in \cite{QT-FRW} (later developed in \cite{QT-BI}, and applied in \cite{QT-pert} to the study of primordial quantum perturbations and possible observable effects in the CMB), where the authors considered the  quantum Klein-Gordon test field propagating on a quantum space-time of the cosmological type in the background. The goal of the works was to derive quantum matter test fields and to ''probe'' with them the quantum nature of the underlying geometry. As a result, quantum equations of  quantum test fields propagating in  quantum space-times were derived. Taking the classical limit in the gravitational degrees of freedom (but leaving the quantum matter field still quantum) gives the standard equations of QFT in a certain classical space-time, say $g_{\rm class}$, in the background. Moreover, it was formulated another, quite natural {\it semi}-classical limit, sensitive to the quantum corrections coming from the quantum nature of the gravitational field \cite{QT-FRW}. In this limit the equations satisfied by the quantum matter test field on the quantum background again take the form of the equations of QFT, however  in a different,  semi-classical space-time. This semi-classical metric differs from $g_{\rm class}$ by corrections depending on quantum fluctuations of the quantum geometry operators, 
therefore it was called quite accurately in \cite{QT-pert} a ''dressed metric'' 
\begin{align} \label{dress-class}
g_{\rm dress}\ =\ g_{\rm class} + O(\hbar).
\end{align}
The dressed metric is a new classical metric\footnote{
It should be clear that $g_{\rm dress}$ is classical, in the sense that it is a metric tensor whose coefficients are functions, not  operators. However, this metric contains quantum corrections and in general  does not satisfy classical Einstein equations.
}
''felt'' by the matter field. Thus, it became clear that effects of the interaction between quantum matter (in the test field approximation) and quantum geometry could be understood by comparing the dressed metric with the expected classical metric. In \cite{QT-FRW}, the dressed metric was computed for a test massless  K-G field in a isotropic quantum space-time of the FRW type. It was observed that the metric was independent of the momentum $\vec{k}$ of considered mode. Therefore, in \cite{QT-BI}  the procedure was applied to the same quantum field on a Bianchi I space-time, with the hope that in this case the dressed metric could depend on the mode's energy and direction of propagation: if that were the case, quanta of different energies would move at different speeds (because they would ''feel'' different dressed metrics), and thus a violation of Lorentz symmetry would be observed. The dressed metric for modes of the massless K-G field was computed: it turned out that such a violation is not present, since the dressed metric is the same for all modes of the matter field (and this result also applies to the subcase case of FRW quantum space-time).  However, a major limitation of that work was present, namely, we were not able to identify the dressed metric in the case of a \emph{massive} K-G field. {The aim of the present paper is to solve this problem, presenting a possible generalization to treat a quantum test massive K-G field on an isotropic quantum space-time.}

Below, we review the concept of dressed metric. Next we calculate the dressed metric for the  quantum test massive K-G field on an isotropic quantum  FRW  space-time. The dressed metric we find is not any longer isotropic, in fact of the Bianchi I type. {In principle, this seems to lead to an emergent Lorentz-violation, but we must be careful: while the isotropy is broken for a given mode, from the point of view of an external observer the behaviour of modes does not depend on their direction. Further work is necessary to understand what kind of measurable effect those dressed metrics can produce. On the techinical side,} our derivation  is quite general and it is independent of a specific choice of the quantum model of space-time: it may well be the original WdW approach, it may be LQC.  For the sake of clarity, we refer to the  LQC models. The same is true for the matter content of  the background quantum space-time and the  ''choice of time'': our result is mostly independent of those, but to be explicit we will use the ''irrotational dust'' introduced recently in \cite{irrD} (see \cite{dust} for further reference).

\section{Classical Theory and Quantization}
\label{CLASS}

We consider the theory described by the action
\begin{align} \label{totalaction}
S = \int d^4x \sqrt{-g} \left[\frac{1}{8 \pi G} R - \frac{1}{2} M \left(g^{\mu \nu} \partial_\mu T \partial_\nu T + 1\right)\right] + S_M
\end{align}
The first term is the usual Hilbert action for geometry, the second term describes \emph{irrotational dust}, and $S_M$ stands for other forms of matter, which eventually will be the Klein-Gordon field. The explicit choice of irrotational dust as part of the matter content is useful to carry out the deparametrization of the theory with respect to $T$. In other words, following \cite{irrD}, we will choose $T$ to represent the physical time.

Since we are interested in the FRW sector of the theory, we consider the symmetry-reduced class of metrics
\begin{align} \label{FRWmetric}
g_{\mu \nu} = -dt^2 + a(t)^2 \left(dx^2 + dy^2 + dz^2\right)
\end{align}
and the dust field is homogeneous on the spatial slices (that is, $\partial_i T = 0$ for $i = 1, 2, 3$). The canonical analysis of the system produces a kinematical phase space $\Gamma_\text{kin}$ coordinatized by the degrees of freedom of the geometry and the matter. Specifically, the momentum conjugate to the dust configuration variable $T$ is given by
\begin{align} \label{dustmomentum}
p_T = a^3 M \dot{T}
\end{align}
The dynamics is completely constrained by the only constraint that survives the symmetry-reduction: the homogeneous part of the total Hamiltonian constraint
\begin{align} \label{hamconstraint1}
C = \int d^3x H
\end{align}
where
\begin{align} \label{hamconstraint2}
H = H_G + H_M + p_T
\end{align}
Here, $H_G$ and $H_M$ are respectively the geometry part and the matter part of the Hamiltonian. Proceeding with Dirac quantization of constrained theories, we define the kinematical Hilbert space as $\mathcal{H}_\text{kin} = \mathcal{H}_G \otimes \mathcal{H}_M \otimes L^2(\mathbb{R}, dT)$, where for now the geometric and matter Hilbert spaces remain unspecified. Formally, the quantum operator on $\mathcal{H}_\text{kin}$ corresponding to $H$ is then
\begin{align} \label{totalhamop}
\widehat{H} = \widehat{H}_G + \widehat{H}_M - i \hbar \partial_T
\end{align}
so physical states $\Psi \in Ker \widehat{H}$ are those $\Psi \in \mathcal{H}_\text{kin}$ such that
\begin{align} \label{implicitschroedinger}
i \hbar \partial_T \Psi(v, q_M, T) = \left[\widehat{H}_G + \widehat{H}_M\right] \Psi(v, q_M, T)
\end{align}
where $v$ is the variable that parametrizes the spectrum of some distingushed operator in  $\mathcal{H}_G$ (see next section) and $q_M$ denotes the collecion of matter variables which coordinatize joint spectrum of suitable set of distingushed operators in $\mathcal{H}_M$.

The form of $\widehat{H}_G$ (and its well-definiteness) depends on the specific quantum theory of gravity, whereas $\widehat{H}_M$ involves both geometrical and matter operators (unless the matter is quantized in a background-independent way). For the sake of clarity, we will consider the LQC quantization (also known as ''polymer quantization'') of the geometrical part, presented in the next section. However, it is important to observe that the results we will find do not rely on this specific choice, and can be repeated for any other proposal.

\section{Polymer Quantum FRW Space-Time}
\label{QS}

Let us focus on the geometric sector $\Gamma_G$ of the phase space. In the symmetry-reduced model of FRW type, $\Gamma_G$ is coordinatized by the scale factor $a$ and its conjugate momentum $p_a$, satisfying the Poisson relation $\{a, p_a\} = 1$. $a$ is positive, so to extend the topology of $\Gamma_G$ to $\mathbb{R}^2$ we perform a canonical transformation to a new set of variables: the \emph{oriented volume} $v := a^3/\alpha$ (where $\alpha = 2 \pi \gamma \sqrt{\Delta} \ell_\text{Pl}^2$, with $\gamma$ is Barbero-Immirzi parameter and $\Delta$ is the so-called ''area gap'' given by $\Delta = 4 \sqrt{3} \pi \gamma \ell_\text{Pl}^2$)
and its conjugate momentum $b$, satisfying $\{v, b\} = 2$. With respect to these variables, the gravitational Hamiltonian is
\begin{align} \label{gravityham}
H_G = \frac{3 \pi G}{2\alpha} b^2 |v|
\end{align}
At the quantum level, polymer representation of the Poisson albebra of $v$ and $b$ is characterized by the Hilbert space $\mathcal{H}_G = L^2(\bar{\mathbb{R}}, d\mu_\text{Bohr})$, where $\bar{\mathbb{R}}$ is the Bohr compactification of the real line and $d\mu_\text{Bohr}$ the Haar measure on it \cite{mathC}. On this Hilbert space it is defined the action of operators $\widehat{v}$ and of $\widehat{N} := \widehat{\exp(ib/2)}$ (the exponentiated version of $b$, since due to discreteness of space the infinitesimal action is not available):
\begin{align} \label{fundamentaloperators}
\widehat{v} | v \rangle = v | v \rangle, \ \ \ \ \ \widehat{N} | v \rangle = | v + 1 \rangle
\end{align}
where $\{| v \rangle\}$ is the basis of eigenstates of $\widehat{v}$, satisfying orthonormality with respect to Kronecher delta $\langle v | v' \rangle = \delta_{v, v'}$. Using a natural symmetric ordering, the gravitational Hamiltonian is implemented as an operator on $\mathcal{H}_G$:
\begin{align} \label{gravityop}
\widehat{H}_G = -\frac{3 \pi G}{8\alpha} \sqrt{|\widehat{v}|} \left(\widehat{N} - \widehat{N}^{-1}\right)^2 \sqrt{|\widehat{v}|}
\end{align}
The properties of this operator are studied in \cite{hamKL}, but for our purposes it is sufficient to say that it well-defined and essentially self-adjoint.

\section{QFT on Quantum Space-time}
\label{QFT}

We now need to choose the matter part of the system. We will consider a scalar K-G field $\phi$ with action
\begin{align} \label{matteraction}
S_M = \frac{1}{2} \int d^4x \sqrt{-g} \left(g^{\mu \nu} \partial_\mu \phi \partial_\nu \phi - m^2 \phi^2\right)
\end{align}
Plugging the ADM metric with lapse $N = 1$ in this, and performing canonical analysis, we obtain the matter Hamiltonian:
\begin{align} \label{matterhamfun}
H_M = \sum_{\vec{k} \in \mathcal{L}} H_{\vec{k}} = \frac{1}{2 a^3} \sum_{\vec{k} \in \mathcal{L}} \left[p_{\vec{k}}^2 + \left(a^4 |\vec{k}|^2 + a^6 m^2\right) q_{\vec{k}}^2\right]
\end{align}
where $\mathcal{L}$ is the three-dimensional lattice spanned by $(k_1, k_2, k_3) \in (2\pi \mathbb{Z})^3$. Notice that $H_M$ is nothing but the Hamiltonian of a collection of decoupled harmonic oscillators with a geometry-dependent (and thus time-dependent) frequency
\begin{align} \label{matterfreq}
\omega_{\vec{k}}^2(a) = a^4 |\vec{k}|^2 + a^6 m^2
\end{align}
For our purposes, it is sufficient to consider a single mode $q_{\vec{k}}$, since we are going to disregard any matter back-reaction in the so-called ''test-field approximation'' (see later). However, it should be said that quantization of the full system would require to take into account all modes. In doing so, renormalization of the UV limit is a crucial element: without it, expressions such as equation (\ref{matterhamfun}) are entirely formal. Quantizing such an infinite-dimensional system is not at all straightforward, and leads to the whole topic of QFT in curved space-time. On the other hand, as long as one is interested only in a single mode (or a finite set), quantization is on the line of quantum harmonic oscillator: the Hilbert space of matter is given by $\mathcal{H}_{M} = L_2(\mathbb{R}, dq_{\vec{k}})$, and the dynamical variables $q_{\vec{k}}$ and $p_{\vec{k}}$ are promoted to operators on it, $\widehat{q}_{\vec{k}} = q_{\vec{k}}$ and $\widehat{p}_{\vec{k}} = -i \hbar \partial/\partial q_{\vec{k}}$.

At this quantum level, the dynamics is described by the Schroedinger equation (\ref{implicitschroedinger}):\footnote{
By $\widehat{a^n}$ we formally mean a quantum realization of the phase space function $a^n$. Of course, such a realization depends on the specific quantum theory of cosmology considered, but in order to reproduce the classical limit its expectation value on a semiclassical state should verify $\langle \widehat{a^n} \rangle = \langle \hat{a} \rangle^n$ at leading order.
}
\begin{align} \label{schroedTOT}
i \hbar & \partial_T \Psi(v, q_{\vec{k}}, T) = \notag
\\
& = \left[\widehat{H}_G + \frac{1}{2} \left(\widehat{a^{-3}} \otimes \widehat{p}_{\vec{k}}^2 + \widehat{a^{-3} \omega_{\vec{k}}^2} \otimes \widehat{q}_{\vec{k}}^2\right)\right] \Psi(v, q_{\vec{k}}, T)
\end{align}
where the geometric operator $\widehat{a^{-3} \omega_{\vec{k}}^2}$ is defined from (\ref{matterfreq}) as
\begin{align} \label{OdefO}
\widehat{a^{-3} \omega_{\vec{k}}^2} = \hat{a} |\vec{k}|^2 + \widehat{a^3} m^2
\end{align}

Now, we use the \emph{test-field approximation}, i.e. the fact that the matter back-reaction is disregarded. This translates mathematically by saying that the total state $\Psi(v, q_{\vec{k}}, T)$ decomposes as a simple tensor product
\begin{align} \label{0th-dec}
\Psi(v, q_{\vec{k}}, T) = \Psi_o(v, T) \otimes \psi(q_{\vec{k}}, T)
\end{align}
for any time $T$, where the geometry state $\Psi_o$ obeys the ''unperturbed'' Schroedinger equation $i \hbar \partial_T \Psi_o = \widehat{H}_G \Psi_o$. In other words, the matter part and the gravity part are disentangled, and evolution of the gravity part does not take into account the presence of matter. Plugging (\ref{0th-dec}) in (\ref{schroedTOT}) and projecting on $\Psi_o$ itself, one is left with a Schroedinger equation for matter only:
\begin{align} \label{schroedMAT}
i \hbar \partial_T \psi(q_{\vec{k}}, T) = \frac{1}{2} \left(\langle \widehat{a^{-3}} \rangle_o \widehat{p}_{\vec{k}}^2 + \langle \widehat{a^{-3} \omega_{\vec{k}}^2} \rangle_o \widehat{q}_{\vec{k}}^2\right) \psi(q_{\vec{k}}, T)
\end{align}
where $\langle \widehat{A}(T) \rangle_o := \langle \Psi_o(v, 0) | \widehat{A}(T) | \Psi_o(v, 0) \rangle$ for every geometrical operator $\widehat{A}$ (time-evolution is moved from the state $\Psi_o$ to the operators, $\widehat{A}(T) = e^{i \widehat{H}_G T/\hbar} \widehat{A} e^{-i \widehat{H}_G T/\hbar}$, thereby realising Heisenberg picture for the gravitational sector (or rather the interaction picture from the point of view of the coupling with the K-G field).

\subsection{The concept of Dressed Metric}
\label{massive}

The heart of the new approach to matter on \emph{quantum} space-time, is the observation that equation (\ref{schroedMAT}) is surprisingly similar to the Schroedinger equation for the states of the quantum field $\phi$ on a suitably defined  \emph{classical} space-time.  Let the geometry be classically described by a metric $\bar{g}_{\mu \nu}$ of the FRW form:
\begin{align} \label{EffGEO1}
\bar{g}_{\mu \nu} dx^\mu dx^\nu = -\bar{N}^2 dt^2 + \bar{a}^2 (dx^2 + dy^2 + dz^2)
\end{align}
One can build regulard QFT on such a curved space-time, obtaining for (a single mode $\vec{k}$ of) a scalar field $\phi$ of mass $m$ the following effective Schroedinger equation:
\begin{align} \label{EffSCHROEDINGER1}
i \hbar \partial_t \psi(q_{\vec{k}}, t) = \frac{\bar{N}}{2 \bar{a}^3} \left(\widehat{p}_{\vec{k}}^2 + \bar{\omega}_{\vec{k}}^2 \widehat{q}_{\vec{k}}^2\right) \psi(q_{\vec{k}}, t)
\end{align}
Comparison with (\ref{schroedMAT}) leads the following system of equations:
\begin{align} \label{sys1}
\bar{N}/\bar{a}^3 = \langle \widehat{a^{-3}} \rangle_o, \ \ \bar{N} \bar{a} |\vec{k}|^2 = \langle \hat{a} \rangle_o |\vec{k}|^2, \ \ \bar{N} \bar{a}^3 m^2 = \langle \widehat{a^3} \rangle_o m^2
\end{align}
This is a system of three equations for two unknowns ($\bar{N}$ and $\bar{a}$), and in general it has no solution for $m\not= 0$. However, in the case $m = 0$,  the last equation drops out. Indeed, for a massless K-G field one has a unique solution:
\begin{align} \label{masslessSol}
\bar{N} = [\langle \hat{a} \rangle_o^3 \langle \widehat{a^{-3}} \rangle_o]^{1/4}, \ \ \ \ \ \bar{a} = [\langle \hat{a} \rangle_o/\langle \widehat{a^{-3}} \rangle_o]^{1/4}
\end{align}
We can then rewrite (\ref{EffGEO1}) explicitely:
\begin{align} \label{DressedGEO1}
\bar{g}_{\mu \nu} dx^\mu dx^\nu & = [\langle \hat{a} \rangle_o^3 \langle \widehat{a^{-3}} \rangle_o]^{1/2} \times
\\
& \times \left(-dt^2 + \frac{1}{\langle \hat{a} \rangle_o \langle \widehat{a^{-3}} \rangle_o} (dx^2 + dy^2 + dz^2)\right) \notag
\end{align}
We managed to express the effective metric in terms of mean values of geometrical operators on the quantum state of geometry $\Psi_o$. This object was defined first in \cite{QT-FRW} and next it has been called the \emph{dressed metric} in \cite{QT-pert}, as it represents the effective classical geometry on which the $\vec{k}$-mode of the matter field lives, in the sense that one can describe the evolution of such a mode on quantum geometry in terms of the same mode propagating on the classical dressed space-time $\bar{g}_{\mu \nu}$. The fact that $\bar{g}_{\mu \nu}$ does not show any dependence on $\vec{k}$ is a hint that no symmetry-breaking takes place: all quanta of matter ''feel'' the same effective metric, probe the same ''eigenstate'' of geometry. The proof that this is indeed true was given via dispersion relation analysis in \cite{QT-BI} (in the more general case of Bianchi I quantum geometry). Someone could thing that this result is trivial, because the back-reaction of the field on the geometry was disregarded. To argue that the result is not trivial, consider the following two points: (i) The dressed metric (\ref{DressedGEO1}) is \emph{not} the classical limit  metric $\tilde{g}_{\mu \nu} dx^\mu dx^\nu = -dt^2 + \langle \hat{a} \rangle_o^2 (dx^2 + dy^2 + dz^2)$; (ii) The back-reaction features  also for the purely  classical theory, nonetheless nobody expects that that the classical back-reaction  generates a Lorentz violation. So back-reaction is not what we want to study.\footnote{
Moreover, it does not seem consistent to take into account only the back-reaction of a single mode: one should indeed consider all modes.
}
What we want to study is the effect of the quantum nature of the geometry on the \emph{test} quantum matter field.  The reader will see in a moment that, in the case of massive K-G field, this effect is emergent isotropy breaking.

\subsection{Massive case}
\label{massive}

We saw above that in the case of the massive K-G field, we could not find a dressed metric. However, our ansatz (\ref{EffGEO1}) for $\bar{g}_{\mu \nu}$ had a drawback: the  assumption of space isotropy. The isotropy seemed natural, because the underlying quantum space-time is isotropic (in the sense that it is obtained by the quantization of isotropic FRW metric). {But in the massive case this guiding principle does not work. Therefore, we have to improve it by requiring that a dressed metric has the same symmetries as the system (but not necessarily more). The full system under consideration consists of the homogeneous isotropic quantum 3-geometries \emph{and} a quantum mode with momentum $\vec{k}$, so a direction is distinguished: hence, the symmetries are those elements of the symmetry group of the quantum 3-geometry which preserve a fixed direction of the momentum $\vec{k}$. It is therefore resonable to enlarge the number of degrees of freedom that describe $\bar{g}_{\mu \nu}$, and consider the Bianchi type I metrics with the suitable symmetry.} Consider first the most general Bianchi I metric:
\begin{align} \label{EffGEO2}
\bar{g}_{\mu \nu} dx^\mu dx^\nu = -\bar{N}^2 dt^2 + \sum_i \bar{a}_i^2 (dx^i)^2
\end{align}
The regular QFT on this curved space-time produces the following  Schroedinger equation:
\begin{align} \label{EffSCHROEDINGER2}
i \hbar \partial_t \psi(q_{\vec{k}}, t) = \frac{\bar{N}}{2 \bar{a}_1 \bar{a}_2 \bar{a}_3} \left(\widehat{p}_{\vec{k}}^2 + \bar{\omega}_{\vec{k}}^2 \widehat{q}_{\vec{k}}^2\right) \psi(q_{\vec{k}}, t)
\end{align}
where
\begin{align} \label{newFreq}
\bar{\omega}_{\vec{k}}^2 = (k_1 \bar{a}_2 \bar{a}_3)^2 + (k_2 \bar{a}_3 \bar{a}_1)^2 + (k_3 \bar{a}_1 \bar{a}_2)^2 + (\bar{a}_1 \bar{a}_2 \bar{a}_3)^2 m^2
\end{align}
Again, comparison with (\ref{schroedMAT}) gives a system of equations:
\begin{align}
\label{sys2a} \frac{\bar{N}}{\bar{a}_1 \bar{a}_2 \bar{a}_3} = \langle \widehat{a^{-3}} \rangle_o, \ \ \ \ \ \bar{N} \bar{a}_1 \bar{a}_2 \bar{a}_3 = \langle \widehat{a^3} \rangle_o
\\
\label{sys2b} \bar{N} \frac{\bar{a}_2 \bar{a}_3}{\bar{a}_1} = \bar{N} \frac{\bar{a}_3 \bar{a}_1}{\bar{a}_2} = \bar{N} \frac{\bar{a}_1 \bar{a}_2}{\bar{a}_3} = \langle \hat{a} \rangle_o
\end{align}
Now, in the generic case of  $k_1,k_2,k_3\not= 0$ there are five equations for four unknowns, so the system is overcomplete and has no solution. Thus, the generic case is excluded. This is good, because a dressed metric should have the symmetry group of the isotropic homogeneous plane orthogonal to the fixed direction $\vec{k}$. In other words, the metrics we are considering are characterised by the diagonal directions such that promoted for the axis of the coordinate system make $\vec{k} = (0, 0, k_3)$.
In this case,  the only equation of (\ref{sys2b}) that does not drop is $\bar{N} \bar{a}_1 \bar{a}_2/\bar{a}_3 = \langle \hat{a} \rangle_o$, which together with the second of (\ref{sys2a}) gives $\bar{a}_3 = \sqrt{\langle \widehat{a^3} \rangle_o/\langle \hat{a} \rangle_o}$. The remaining two variables must satisfy $\bar{a}_1 \bar{a}_2 = \sqrt{\langle \widehat{a^3} \rangle_o \langle \hat{a} \rangle_o}/\bar{N}$, {so there are infinitely many solutions. However, the one that has the required symmetry is that with $\bar{a}_1 = \bar{a}_2$.}\footnote{
{It makes sense that the particle does not distinguish between two metrics differing only in the scale factors $\bar{a}_1$ and $\bar{a}_2$, since these are orthogonal to the direction of propagation, and therefore interest a part of space-time that is not accessible to the ''probing power'' of the particle. In this sense, one should think of the dressed metric not as giving information about the whole space-time geometry, but only about the part which is relevant for the mode.}
}
This fixes a unique solution
\begin{align}
\label{masslessSola} \bar{N} = \sqrt{\langle \widehat{a^3} \rangle_o \langle \widehat{a^{-3}} \rangle_o}, \ \ \ \ \ \bar{a}_3 = \sqrt{\langle \widehat{a^3} \rangle_o/\langle \hat{a} \rangle_o}
\\
\label{masslessSolb} \bar{a}_1 = \bar{a}_2 = \left[\langle \hat{a} \rangle_o\right/\langle \widehat{a^{-3}} \rangle_o]^{1/4}
\end{align}
The dressed metric is thus of the Bianchi I type (\ref{EffGEO2}), despite the underlying quantum geometry is FRW. At first sight, this anisotropy might not be so surprising: after all, we are considering a specific particle, moving in a specific direction (namely, $\vec{k}$), and thus it is natural to expect that the back-reaction should have a direction-dependent effect on the geometry. This is certainly true, but one has to remember that in our analysis we completely disregarded any back-reaction of matter on geometry. In other words, the space-time does not know anything about the existence of the propagating particle. Indeed, what is deformed is the effective metric \emph{felt} by the particle, not the space-time itself: an external observer would still measure a space-time of the FRW type.

In the dressed metric, a certain degree of symmetry remains: the scale factors in the directions orthogonal to the direction of propagation of the particle are unchanged with respect to the massless case (i.e., $\bar{a}_1 = \bar{a}_2 = \bar{a}$). What changes is the scale factor in the direction $\vec{k}$ (here the $z$-direction) and the lapse function. In particular, if we denote the FRW lapse function in (\ref{masslessSol}) by $\bar{N}_o$, then we find
\begin{align} \label{FRW-B1}
\bar{N} = \alpha \bar{N}_o, \ \ \ \ \ \bar{a}_3 = \alpha \bar{a}
\end{align}
where
\begin{align} \label{FRW-B2}
\alpha := \left[\frac{\langle \widehat{a^3} \rangle_o^2 \langle \widehat{a^{-3}} \rangle_o}{\langle \hat{a} \rangle_o^3}\right]^{1/4}
\end{align}
As expected, $\alpha$ reduces to $1$ if we can write $\langle \widehat{a^3} \rangle_o = \langle \hat{a} \rangle_o^3$ and $\langle \widehat{a^{-3}} \rangle_o = \langle \hat{a} \rangle_o^{-3}$ (as would happen if the geometry were completely classical), thus recovering the FRW case (more precisely, we would get the classical limit metric $\bar{a}_1 = \bar{a}_2 = \bar{a}_3 = \langle \hat{a} \rangle_o$). This proves that the deformation of the symmetry is an effect of the quantum nature of the geometry, confirming that at $0$th order in the back-reaction there is no classical effect on the geometry.

It is possible to expand $\alpha = 1 + \delta \alpha$. To do this, define the quantities $\delta A := (\langle \widehat{a^3} \rangle_o - \langle \hat{a} \rangle_o^3)/\langle \hat{a} \rangle_o^3$ and $\delta B := (\langle \widehat{a^{-3}} \rangle_o - \langle \hat{a} \rangle_o^{-3})/\langle \hat{a} \rangle_o^{-3}$, which measure the ''non-classicality'' of the geometry in a way similar to the usual variance. Then, we can write
\begin{align} \label{???3}
\alpha & = \left[(1 + 2 \delta A + \delta A^2) (1 + \delta B)\right]^{1/4} \approx \notag
\\
& \approx 1 + \frac{1}{2} \delta A + \frac{1}{4} \delta B
\end{align}
having retained only the first order corrections in $\delta A$ and $\delta B$. Notice that for large times the quantum nature of geometry can be neglected \cite{L-FRW3}, so one can argue that $\delta A$ and $\delta B$ vanish in the classical gravity regime: in other words, a deviation from the FRW metric is present only at early times, i.e. in the vicinity of the bounce.

\section{Conclusions}
\label{conclusions}

{We  have studied quantum test fields on a quantum homogeneous and isotropic universe. The quantum universe is described by a quantum state of  geometry $a$ (the scale factor) coupled to a background homogeneous irrotational dust field $T$. We did not assume much more about the quantum cosmological model: it may be a LQC model, but it can also be an old-fashioned WdW model. Qualitative results are not sensitive on those details.} {Also the choice of physical time is not particularly relevant here.}\footnote{
{For the usual choice of time (namely, a homogeneous massless K-G field $T$ -- see \cite{scal1, scal2} for further reference), one obtains a Schroedinger-like equation of the form $i \hbar \partial_T \Psi = \sqrt{\widehat{H}_G^2 - 2 \widehat{H}_M} \Psi$. In this case, it is necessary to use an operator expansion, which produces an approximate dynamical equation $i \hbar \partial_T \Psi = [\widehat{H}_G - \widehat{H}_G^{-1/2} \widehat{H}_M \widehat{H}_G^{-1/2}] \Psi$. A part from this modification, the argument then proceeds as above.}
}

{First, we recalled that in this quantum universe all the modes of a test \emph{massless} K-G field $\phi$  behave as if they were in a classical spacetime
\begin{equation}
\bar{g}_o\ =\ -\bar{N}_o^2dT^2 + \bar{a}^2(dx^2+dy^2+dz^2)
\end{equation}
where the metric coefficients $\bar{N}_o$ and $\bar{a}$ are defined in (\ref{masslessSol}) as the products of expectation values of suitable operators (which evolve with $T$). The fact that $\bar{N}_o$ and $\bar{a}$ are not simply the expenctation values of the associated operators makes the effective classical metric $\bar{g}_o$ ``dressed''.}

{Then, we considered a test \emph{massive} K-G field on the same quantum universe. We found that, given a direction (unitary vector) $\vec{n}$ in the space of $3$-momenta, all the modes $\psi(q_{k\vec{n}})$ of the test K-G field propagate in the quantum universe as if they were feeling an effective classical spacetime with metric tensor 
\begin{equation} \label{metricF}
\bar{g}_{\vec{n}}\ =\ \bar{g}_o\ +\ (\alpha^2 - 1)\left(-\bar{N}_o^2dT^2 + \bar{a}^2 (n_adx^a)^2\right)
\end{equation}
where $\alpha$ is given in (\ref{FRW-B2}) as product of expectation values of $T$-dependent operators. This new dressed metric is homogeneous on the surfaces $T = const$, but is no longer isotropic on them. The generalization of the concept of dressed metric to the massive case is the main result of the present work.}

{A theoretical application of such result is in the construction of a QFT for massive modes on quantum cosmological spacetime. For every direction $\vec{n}$ in the space of momenta, the Hilbert space of quantum states of the modes $\hat{q}_{k\vec{n}}$ is given by the Fock space for the quantum modes in the classical dressed spacetime $\bar{g}_{\vec{n}}$. The total Hilbert space is obtained by combining the states defined for each $\vec{n}$.}

{Observational consequences are more subtle, but we shall present here our understanding of the physics behind the emergent dressed metric (\ref{metricF}). If the only physical information we can extract from the metric comes from the dispersion relation, then there is little chance to observe any anisotropy. Indeed, even if the metric (\ref{metricF}) depends explicitely on the mode $\vec{k}$ (via its direction $\vec{n}$), given a certain mode there is always a system of coordinates such that the metric is of the diagonal form, with deformation along the $z$-direction. Now, since the dispersion relation is a scalar (the so-called \emph{mass-shell}), it can be computed in any coordinate system, so in particular in the one adapted to $\vec{k}$. Doing this for each single mode, one always finds the same dispersion relation, losing any dependence on the direction. Nevertheless, the fact remains that the scale factor to be used (the one oriented along $\vec{n}$) is $\alpha \bar{a}$ rather than $\bar{a}$, which means that} {a massive particle will behave as if there was an additional force acting on it, which does not have origin in the metric $\bar{g}_{o}$.} {This is an effect to be considered, despite not being $\vec{k}$-dependent.}

{Here we only analyzed the dispersion relation, and the conclusion is that \emph{no isotropy-breaking effects are present}. However, more sophisticated observations could detect interesting behaviours. An example is the interaction between two non-parallel modes. In that case, we cannot take into account the direction of propagation of each mode separately, but rather the direction of one mode ''seen'' by the other. That is to say: no coordinate transformation can diagonalize both dressed metrics, $\bar{g}_{\vec{n}}$ and $\bar{g}_{\vec{n}'}$, at the same time. How this fact could affect the interaction between the two modes is unclear to us, and -- plagiarizing a man wiser than us -- \emph{hypotheses non fingimus}.}

{Whatever observable effects there might be, we should finally underline that the ''dressing'' requires that gravity be strongly non-classical, and hence the deformation takes place only in the vicinity of big bang (or big bounce, in LQC), i.e. in the primordial past. This would explain why nowadays we do not observe any such effect whatsoever. Nevertheless, in studying QFT for massive fields in the very early universe, our dressed metric (\ref{metricF}) should be considered. In principle, it might have played an important role on initial quantum fluctuations, possibly leaving a signature in the cosmic microwave background (CMB) or influencing the formation of structures.}

\section{Acknowledgments}

This work was partially supported by the grant of Polish Ministerstwo Nauki i Szkolnictwa Wy\.{z}szego nr N N202 104838 and by the grant of Polish Narodowe Centrum Nauki nr 2011/02/A/ST2/00300.


\end{document}